\documentstyle[prb,preprint,aps,tighten,graphicx]{revtex}
\begin{document}
\draft

\title{Theoretical Study of Electrical Conduction Through a
Molecule Connected to Metallic Nanocontacts}

\author{Eldon G. Emberly\footnote{e-mail: eemberly@sfu.ca           
Copyright 1998 The American Physical Society} and George Kirczenow}

\address{Department of Physics, Simon Fraser University,
Burnaby, B.C., Canada V5A 1S6}

\date{\today}

\maketitle
\begin{abstract}
We present a theoretical study of electron transport through a
molecule connected to two metallic nanocontacts.  The system
investigated is 1,4 benzene-dithiolate (BDT) chemically bonded
to two Au contacts.  The surface chemistry is modeled by
representing the tips of the Au contacts as two atomic
clusters and treating the molecule-cluster complex as a single
entity in an extended H\"uckel tight binding scheme.  We model
the tips using several different cluster geometries.  An ideal
lead is attached to each cluster, and the lead to lead
transmission is calculated.  The role of the molecule-cluster
interaction in transport is analyzed by using single channel
leads.  We then extend the calculations to multi-channel leads
that are a more realistic model of the tip's environment.
Using the finite-voltage, finite temperature
Landauer formula, we calculate the differential conductance
for the different systems studied. The similarities and
differences between the predictions of the present class of
models and recent experimental work are discussed.
\end{abstract}
\pacs{PACS: 73.40.-c, 73.61.Ph, 73.23.-b}
\section{Introduction}
A molecular wire in its simplest definition consists of a
molecule connected between two reservoirs of electrons.  Such
a system poses many interesting theoretical and experimental
challenges.  It was first suggested in the early '70's by
Aviram and Ratner that such a system should have the ability
to rectify current\cite{Avi74}.  Recent experiments on
molecular wires have included studies of conduction in
molecular thin films\cite{Fisch95,Zhou97}, in self-assembled
monolayers (SAM's) using an
STM\cite{Andres96,Mirkin96,Bumm96,Stipe97,Datta97_2}, and
through a single molecule connected between the tips of a
mechanically controlled break junction\cite{Reed96}.
Theoretical work has also appeared recently on electronic
transport in these mesoscopic
systems\cite{Samant96,Kemp96_1,Mago97,Datta97_1,Ember98_1,Datta97_2}.

The theoretical analysis of conduction through a molecule
bonded to metallic contacts brings together different methods
from chemistry and physics.  The treatment of the molecule
itself is a problem in quantum chemistry.  Many different
techniques exist for the calculation of the electronic
structure of molecular systems.  There is also the matter of
how to treat the interaction between the molecule and the
surface of the metallic reservoir.  This can be done using an
effective interaction via Newns-Anderson\cite{Mujic96} or by
modeling the tips atomistically\cite{Joach96}.  Once these
issues have been addressed it is possible to proceed to the
electron transport problem.  For molecular wires with
nano-meter dimensions, Landauer theory\cite{Lan57,Datta95} is
used which relates the conductance to the electron transmission
probability.

A molecule of current experimental interest as a molecular
wire,\cite{Reed96} and the one studied theoretically in this
article, is 1,4 benzene-dithiol which when attached to two gold
leads becomes 1,4 benzene-dithiolate (BDT). It consists of a
benzene molecule with two sulfur atoms attached, one on either
end of the benzene ring.  The benzene offers delocalized
electrons in the form of $\pi$ orbitals, conducive to electron
transport.  The sulfurs bond effectively to the gold
nanocontacts.  Two major unknowns of the experimental
system\cite{Reed96} are the geometry of the gold contacts and
the nature of the bond between the molecule and these contacts.
This paper attempts to address these important issues.

In our model of the BDT wire we attach the molecule to two gold
clusters.  Thus we consider the interactions between the
molecule and tip surface at an atomistic level.  This system of
clusters+molecule (CMC) is then treated as a single larger
molecule which we model using the extended H\"uckel tight
binding method of quantum chemistry.  This allows us to address
the chemical nature of the interface between the nanocontacts
and molecule by treating the clusters and BDT as an integral
entity.  In this paper we consider gold clusters oriented in
the (100) and (111) directions.  We also examine different
binding schemes for the thiol end-groups of the BDT to these
clusters.

Using Landauer theory and the above tight binding model of the
contact region, we proceed to study electronic transport
through the system by attaching ideal leads to the backs of the
clusters.  We consider two types of leads in this paper.  The
first is a 1D lead consisting of a chain of atoms with just one
conducting electronic mode or channel.  It will be seen that
this allows for a careful examination of how the chemistry of
the CMC system, specifically the hybrid energy states arising
out of mixtures of cluster and molecule orbitals, affects
transport.  The second model for the leads considers them to be
multi-mode or multi-channel.  The multi-channel lead is
constructed by periodically repeating a multi-atom unit cell in
1D.  This is a more realistic model of nanocontacts which in
reality have many different electron modes propagating at a
given energy.

The strength of the coupling between the molecule and the
clusters is a significant factor in the control of electron
transport through the wires.  For single-channel ideal leads,
we find that for strongly coupled systems the highest occupied
and lowest unoccupied molecular orbital (HOMO/LUMO), do not
necessarily control the conduction of electrons.  For this case
transport is mediated by states that are mixtures of those in
the clusters and the molecule.  In our calculations with
multi-mode leads, the same is true, and also the molecule is
found to be very conductive (more so than has been reported
experimentally\cite{Reed96}).  By increasing the bond lengths
between the sulfur and the gold clusters, thereby weakening the
coupling, it is found that the magnitude of the transmission
decreases and that the resonances in the transmission can be
related to energy levels of the isolated molecule.  However, to
reduce the transmission to experimental levels the bond lengths
have to be stretched to unphysical dimensions.

In Sec. II, we describe the scattering and transport theory
used in this work.  It is a method for solving the
tight-binding form of the Schroedinger equation for the $t$
matrix and transmission coefficients.  The electronic structure
of isolated BDT is discussed in Sec. III.  To explore the
effects of coupling in the simplest possible context, the
results of a transmission calculation for BDT connected
directly to ideal single-channel leads are described in
Sec. IV.  The evolution of the energy eigenstates of the system
when Au clusters are attached to the BDT is discussed in
Sec. V.  The calculated transmission and conductance for
various BDT-cluster systems attached to one mode leads is
presented in Sec. VI.  Sec. VII describes the calculated
transmission and conductance for multi-mode leads.  Our
conclusions are presented in Sec. VIII.
\section{Transport Theory}
\subsection{Landauer Formula}
We consider the transport of electrons through a molecular
system by modeling it as a one electron elastic scattering
problem.  The molecule acts as a defect between two metallic
reservoirs of electrons.  An electron incident from the source
lead with an energy $E$, has a transmission probability $T(E)$
to scatter into the drain lead.  By determining the
transmission probability for a range of energies around the
Fermi energy, $\epsilon_{F}$ of the lead, the finite
temperature, finite voltage, Landauer formula can be used to
calculate the transmitted current $I$ as a function of the bias
voltage, $V$, applied between the source (left lead) and drain
(right lead)
\begin{equation}
I(V) = \frac{2e}{h} \int_{-\infty}^{\infty}
dE\:T(E)\left( \frac{1}{\exp[(E-\mu_{s})/kT] + 1} -
\frac{1}{\exp[(E-\mu_{d})/kT]+1} \right)
\end{equation}
The two electro-chemical potentials $\mu_{s}$ and $\mu_{d}$,
refer to the source and drain, respectively.  They are defined
to be, $\mu_{s} = \epsilon_{F} + eV/2$ and $\mu_{d} =
\epsilon_{F} - eV/2$.  The differential conductance is then
given by the derivative of the current with respect to voltage.

\subsection{Evaluation of the Transmission Matrix}
We find the multi-channel transmission probability $T(E)$ by
solving the Schroedinger equation directly for the scattered
wavefunctions. This is done by setting up a set of equations
involving the transmission and reflection matrices for the
modes of the leads that are coupled to the molecule.  We treat
the Hamiltonian using the tight-binding approximation in a
non-orthogonal atomic basis set.  The formalism we use differs
from other tight-binding methods involving Green's function
techniques\cite{Samant96} and transfer matrix
methods.\cite{Sautet88} For the one-electron tight-binding
Hamiltonians used it yields exact results.

\begin{figure}[ht]
\includegraphics[bb= 20 200 600 530, width=0.75\textwidth,clip]{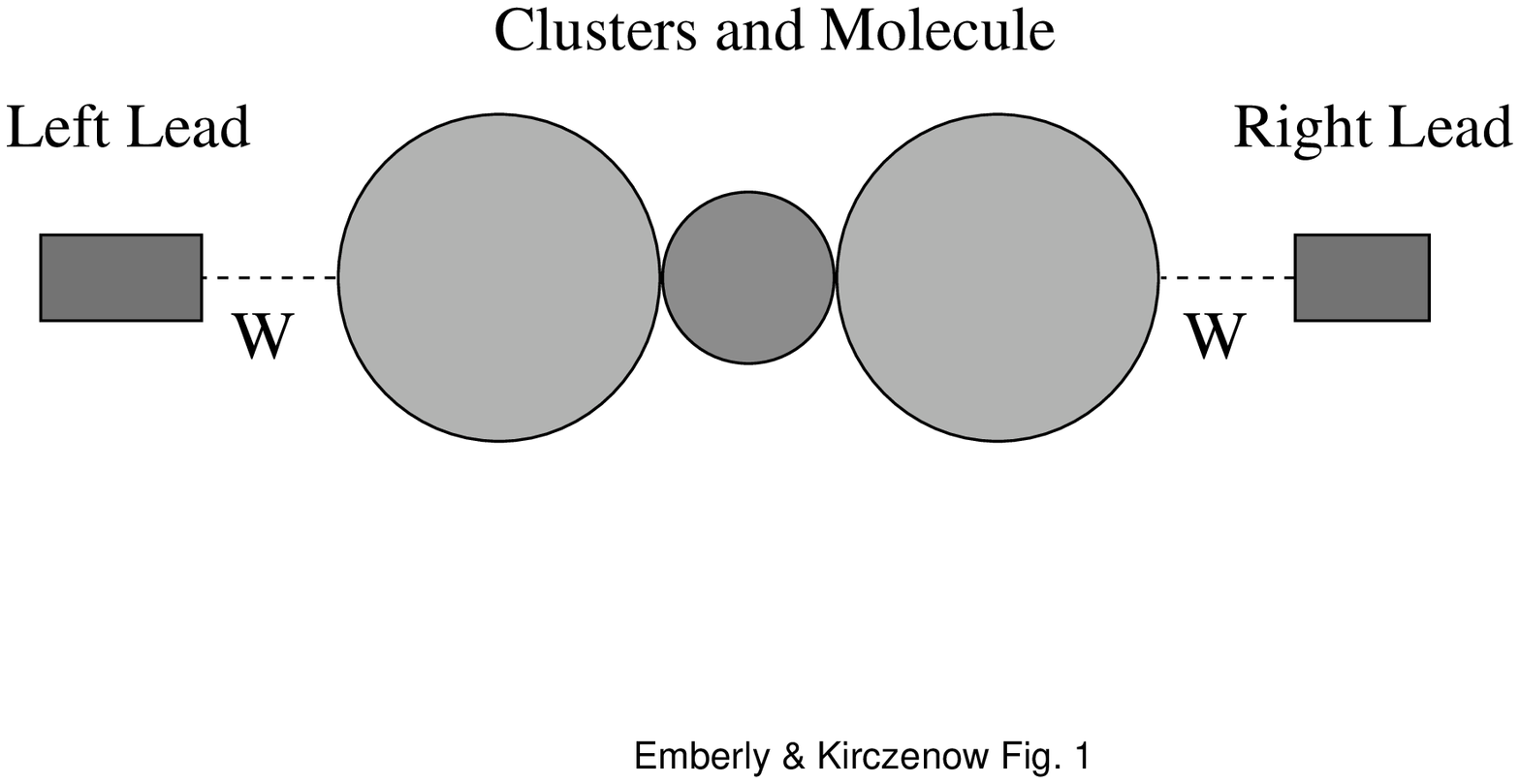}
\caption{Schematic diagram of the three non-interacting
systems: left lead, CMC, and right lead.  The systems are
coupled by the potential $W$.}
\label{fig:1}
\end{figure}

We start with Schroedinger's equation, $H|\Psi^\alpha\rangle =
E|\Psi^\alpha\rangle$, where $H = H_o + W$ is the Hamiltonian
for the entire coupled system composed of the left lead,
clusters + molecule (CMC) and right lead.  The Hamiltonian
$H_o$ is that for the decoupled system consisting of isolated
left and right leads and the CMC.  The ideal leads are then
coupled to the CMC system via the coupling potential $W$.  A
schematic of the model system is shown in Fig. \ref{fig:1}.
The wavefunction $|\Psi^\alpha\rangle$, which describes an
electron with energy $E$, propagating initially in the
$\alpha^{th}$ mode of the left lead, will be written using LCAO
in a non-orthogonal set $\{|n,j\rangle\}$.  Here $n$ labels the
site (or unit cell) and $j$ labels the atomic orbital on the
site.  $|\Psi^\alpha\rangle$ is expressed in terms of the
transmission and reflection matrices, $t_{\alpha,\alpha'}$ and
$r_{\alpha,\alpha'}$ and has different forms in the left lead
(L), CMC (M), and right lead (R).  The total wavefunction is a
sum of these three, $|\Psi^{\alpha} \rangle =
|\Psi^{\alpha}_{L}\rangle + |\Psi^{\alpha}_{M}\rangle +
|\Psi^{\alpha}_{R}\rangle$, where
\begin{eqnarray}
|\Psi^{\alpha}_{L}\rangle &=& \sum_{n=-\infty}^{n=-1} \sum_j
 (c_{j}^{\alpha} e^{i n y^{\alpha}} + \sum_{\alpha' \epsilon
 L} r_{\alpha',\alpha} c_{j}^{\alpha'} e^{i n
 y^{\alpha'}})|n,j\rangle \\
|\Psi^{\alpha}_{M}\rangle &=&
 \sum_j a_{j}^{\alpha} |0,j\rangle \\
|\Psi^{\alpha}_{R}\rangle &=& \sum_{n=1}^{n=\infty} \sum_j
 \sum_{\alpha' \epsilon R} t_{\alpha,\alpha'} d_{j}^{\alpha'}
 e^{i n y^{\alpha'}} |n,j\rangle
\end{eqnarray}
The forms for the wavefunction in the left and right leads can
be interpreted in the following way.
$|\Psi^{\alpha}_{L}\rangle$ is composed of a rightward
propagating Bloch wave along with reflected leftward
propagating and decaying evanescent modes in the left lead.
For the right lead, $|\Psi^{\alpha}_{R}\rangle$, is a sum over
the transmitted rightward propagating and decaying evanescent
modes in the right lead.  All modes in both leads have energy
$E$.  Note that in the leads the site index $n$ labels the unit
cells of the leads and the sum over $j$ is over all orbitals in
a multi-atom unit cell.  The CMC is assigned the site label
$n=0$ and there $j$ runs over all of the atomic orbitals of the
CMC.  The Bloch coefficients, $c^{\alpha}$ and $d^{\alpha}$,
for the left and right leads respectively and the reduced
wavenumbers $y^{\alpha}$ are calculated for the various modes
using a transfer matrix method outlined in the appendix.

Using Schroedinger's equation and applying the bra $\langle
n,i|$, the following system of linear equations is arrived at
for the unknown quantities, $r_{\alpha',\alpha}$, $a_j$, and
$t_{\alpha',\alpha}$
\begin{equation}
\sum_{\alpha' \epsilon L} r_{\alpha',\alpha}
\sum_{m=-\infty,j}^{m=-1} A_{n,m}^{i,j} c_{j}^{\alpha'}
e^{imy^{\alpha'}} + \sum_j A_{n,0}^{i,j} a^{\alpha}_{j} +
\sum_{\alpha' \epsilon R} t_{\alpha',\alpha}
\sum_{m=1,j}^{m=\infty} A_{n,m}^{i,j} d_{j}^{\alpha'}
e^{imy^{\alpha'}} = - \sum_{m=-\infty,j}^{m=-1} A_{n,m}^{i,j}
c_{j}^{\alpha} e^{imy^{\alpha}}  \label{eq:lineqns}
\end{equation}
where $A_{n,m}^{i,j} = H_{n,m}^{i,j} - E S_{n,m}^{i,j}$ with
the Hamiltonian matrix, $H_{n,m}^{i,j} = \langle
n,i|H|m,j\rangle$ and the overlap matrix, $S_{n,m}^{i,j} =
\langle n,i|m,j \rangle$, being introduced explicitly.  The
overlap matrix is defined as the overlap between orbitals $i$
and $j$ on sites $n$ and $m$ respectively.  This incorporates
the non-orthogonal basis set into the theory.  The Hamiltonian
matrix consists of a sum of the free Hamiltonian matrix,
$(H_o)_{n,m}^{i,j}$, and the coupling matrix, $W_{n,m}^{i,j}$.

We now proceed to define the relevant matrices for our model.
As with the wavefunctions, the free Hamiltonian, $H_{o}$, is
partitioned into the three systems, (L), (R) and (M). We
assume that the atomic orbitals on each lead are mutually
orthogonal and only hopping between nearest neighbour cells is
considered.  Thus the Hamiltonian matrix for the leads is
composed of a matrix $\epsilon$ consisting of (atomic) site
and intra-cell hopping energies and a matrix $\beta$ of all
the inter-cell hopping energies between atoms of nearest
neighbour cells.  The Hamiltonian elements for the CMC are
taken to be extended H\"uckel matrix elements between all
orbitals on the CMC.  The extended H\"uckel method is an
empirical LCAO quantum chemistry method that provides a
reasonable approximation to energies for molecular systems
such as those being studied in this paper.  More sophisticated
ab-initio quantum chemistry methods can also be used to
evaluated these matrix elements.

We also use extended H\"uckel to estimate the interaction
between the leads and the CMC.  The coupling matrix, $W$,
is taken to consist of the H\"uckel matrix elements between
all of the orbitals in the unit cells directly adjacent to the
CMC and all orbitals of the CMC.  All other matrix elements
are assumed to be zero.

Some assumptions are also made about the overlap matrix
$S_{n,m}^{i,j}$.  As was mentioned above, the orbitals on the
leads are assumed to be orthogonal to one another.  Also, the
overlap between orbitals on the lead and orbitals on the CMC
will be assumed zero.  However, the orbitals on the CMC will
not be assumed to be orthogonal to each other.

With the matrices that enter Eq. (\ref{eq:lineqns}) thus
defined we solve Eq. (\ref{eq:lineqns}) numerically for the
reflection and transmission coefficients $r_{\alpha',\alpha}$
and $t_{\alpha',\alpha}$ for each rightward propagating mode
$\alpha$ at energy $E$ in the left lead.  The total
transmission and reflection are then given by
\begin{equation}
T(E) = \sum_{{\alpha \epsilon L}} \sum_{{\alpha'
\epsilon R}} \frac{v^{\alpha'}}{v^{\alpha}}
|t_{\alpha',\alpha}|^{2}
\end{equation}
\begin{equation}
R(E) = \sum_{{\alpha \epsilon L}} \sum_{{\alpha'
\epsilon L}} \frac{v^{\alpha'}}{v^{\alpha}}
|r_{\alpha',\alpha}|^{2}
\end{equation}
where $v^{\alpha}$ is the velocity of the electron in the
$\alpha^{th}$ rightward propagating mode in the left lead.  The
sum over $\alpha'$ in the expression for $T$ is over the
rightward propagating modes at energy $E$ in the right lead.
For $R$ the sum over $\alpha'$ is over the leftward propagating
modes in the left lead.

\section{Electronic Structure of BDT}
The molecule studied in this paper is 1,4 benzene-dithiolate
(BDT).  The ring of $\pi$ conjugated carbon atoms provides
delocalized electrons beneficial for conduction.  The sulfur
atoms on either end of the molecule bind to the two metallic
leads.  The nature and effects of this binding will be explored
in the sections to follow.  However, before examining the
coupled system it is important to consider the electronic
structure of the isolated molecule.

\begin{figure}[ht]
\includegraphics[bb= 40 140 550 680 ,width=0.75\textwidth,clip]{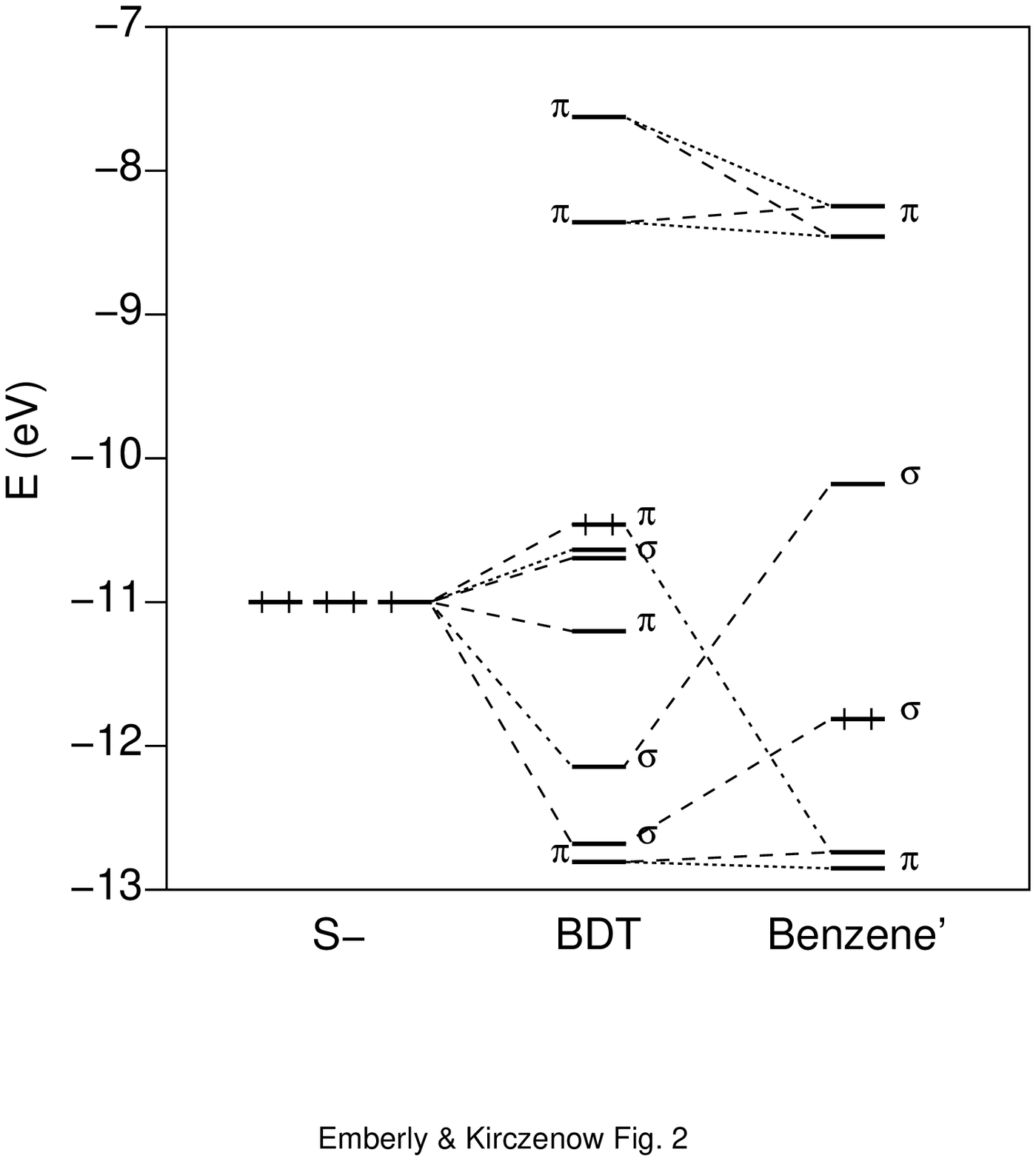}
\caption{Energy level diagram for BDT ion.  Left diagram shows the p
orbitals for one of the sulfur ions (S) (Not shown is the sulfur S
level, which is lower in energy).  Right diagram is for benzene with
the 1,4 H's removed.  The dashed lines connecting the diagrams
indicate where the levels of a given segment get mixed into the BDT
ion.  The HOMO's are identified as the levels with the vertical line
segments crossing the energy level.}
\label{fig:2}
\end{figure}

In Fig. \ref{fig:2} the electronic structure of BDT is shown in
the middle diagram.  The energy levels were calculated using
extended H\"{u}ckel.  The geometry of the molecule was taken to
be that of benzene with the 1,4 hydrogens replaced with sulfur.
The sulfur-carbon bond was taken to be 1.795 Angstroms.  It
should be noted that the molecule is technically an ion since
the sulfurs each have an extra electron, given up by the
hydrogens in changing from a thiol to a thiolate.  The BDT is
similar to simple benzene in that its HOMO and LUMO are both
$\pi$ like orbitals.  To the right of BDT in the figure are the
energy levels for benzene with the H's at positions 1 and 4
removed (labeled Benzene').  The HOMO-LUMO gap for BDT is
smaller than that of benzene.  For benzene (not benzene') the
HOMO is at -12.814 eV with a LUMO of -8.248 eV.  It is shown in
the diagram where the benzene' levels mix into the BDT.  For
BDT the HOMO is at -10.470 eV and the LUMO is at -8.247 eV.
The character of the BDT HOMO is that of C $\pi$ on the ring
with some S $\pi$ content.  The other levels in the BDT
spectrum, as marked on the diagram, also contain C $\pi$ and C
$\sigma$ bonding states.  The $\sigma$ states are due to the
two states in the benzene' spectrum that occur within the
HOMO-LUMO gap of benzene.  The figure also shows where the 2p
orbitals of the sulfur (on the left in the diagram) mix into
the molecular orbitals.  The levels around the BDT HOMO contain
significant S content, while the levels around the LUMO have
less S character.  The levels that have S character will be
influenced strongly when the molecule is bonded to the gold
clusters.

\section{Transmission in a Simple Coupled BDT System}
\begin{figure}[ht]
\includegraphics[bb= 0 75 550 680 ,width=0.75\textwidth,clip]{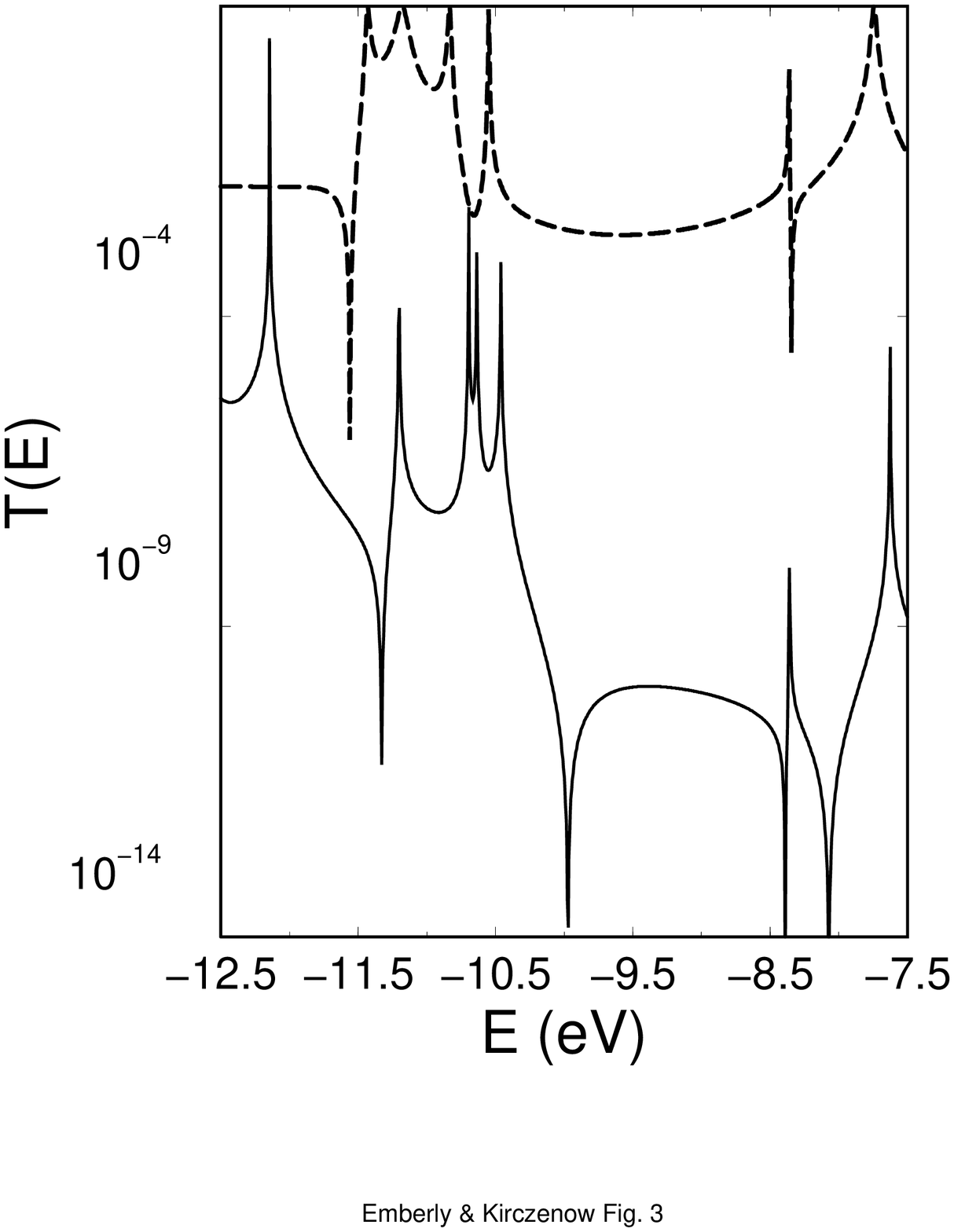}
\caption{Transmission diagrams for BDT connected directly to
two ideal leads.  The solid line corresponds to weak coupling,
while the dashed line is for strong coupling to the molecule.}
\label{fig:3}
\end{figure}

We now proceed to apply the transport theory described in
Sec. II by considering to start with a very simple model
system: a CMC consisting only of a BDT molecule bonded directly
to ideal 1D leads.  The unit cell of the leads has one gold
atom with only 6s orbitals.  Thus there is only one energy band
and only a single mode for the incident electrons.  This allows
us to explore the transmission in the simplest possible
environment.  We use the H\"{u}ckel matrix elements between the
orbitals on the two gold atoms adjacent to the molecule and all
orbitals on the molecule as the matrix elements of $\bf{W}$.
Two cases of coupling will be studied:

The first case is of weak coupling.  This means that the leads
are not well bonded to the molecule with a lead to sulfur
distance greater than 4 Angstroms.  Because of the weak
interaction between the leads and the molecule one would
expect the molecular levels not to be significantly altered by
the presence of the leads.  This is indeed the case and as is
seen in Fig. \ref{fig:3}. The solid curve, which represents
the transmission through the molecule at weak coupling, has
resonances (transmission maxima) at energies close to the
molecular levels of isolated BDT.  The relative strengths of
the different resonances shown appear to be reasonable since
from the energy level diagram of BDT the levels around the
HOMO should be more conductive, due to having more S
character.  With less S content, the levels around the LUMO
are less conductive.

The second case that we consider is typical of more realistic
bonding distances of less than 3 Angstroms.  In this case the
interaction between the lead and the molecule is not weak and
one would expect the molecular levels to be influenced by the
presence of the leads.  The dashed line in the transmission
plot is for this case.  The resonances have shifted and
broadened; some have disappeared.  The nature of the levels
has also changed, for example the resonance at around -8 eV
is now due to a $\sigma$ level rather than a $\pi$ LUMO
level.  This is a first example of the effects of mixing and
much more will be made of this below.

Another interesting phenomenon, is the occurrence of
antiresonances where there is almost perfect reflection of the
incident electron from the molecule.  This arises from
interference between the different molecular energy levels and
from dynamics associated with the non-orthogonality of orbitals
on different sites.\cite{Kemp96_2,Ember98_2} It will also be
seen to occur in the results to follow.

The purpose of the above calculations was to show how even for
a simple model the coupling between the molecule and leads
plays an important role.  For weak coupling, the molecular
resonances survive and the transmission is identifiable with
the molecular energy level structure.  However for strong
coupling the electronic structure is modified significantly,
which leads to resonances which depart from those of the
isolated molecule and are often not related to the isolated
molecular levels in any simple way.  Clearly the chemistry
needs to be considered carefully in studying transport in the
more realistic models to be discussed next.

\section{CMC Models: Different Binding Schemes}
We now proceed to analyze the electronic structure and relevant
chemistry for some models of CMC systems which we believe may
represent the atomic structure of the contact region.  It has
been shown experimentally \cite{Laibin91} that when 1,4
benzene-dithiol reacts with gold, it loses a pair of H ions to
become BDT.  The H ions react with the gold by capturing a
valence electron and the sulfur bonds to the gold.  No studies
have as yet been done to characterize the nature of the
interface between the sulfur and the gold nanowire.  Recent
studies of gold nanowires and break junctions of macroscopic
gold wires have shown them to be different from bulk gold; the
region between two gold contacts in a break junction is
composed of filament structures\cite{Corr97,Kondo97}.  The
reconstructions of the Au atoms for a nanobridge tend to form
hcp structures.  Although it is not known whether either of the
systems described in these reports is characteristic of the
exact nature of the sulfur-gold interface, it seems reasonable
to model the leads with different geometrical configurations
that are consistent with the above findings for nanoscale gold
structures.

Another area of uncertainty is the nature of the bond between
the sulfur and the gold.  An SCF calculation by Sellers {\em
et al.}\cite{Sell93} has yielded calculated bond lengths and
angles for alkanethiolate bonded over gold (100) and (111)
surfaces.  It is possible for the sulfur to bind over either a
hollow site on the surface or directly over a gold atom.  The
binding over a hollow site has been calculated to be
energetically favourable.  Experiments on self-assembled
monolayers (SAM's) tend to support the binding over hollow
sites\cite{Bumm96}.  Recent theoretical studies have also
shown that the hollow site is the favourable binding site for
alkanethiol SAM's\cite{Beard98}.  Similar calculations for
BDT-gold are not available at this time.  Thus in the present
work we study configurations in which the sulfur is bound over
a gold atom as well as those in which it is bound over a
hollow site.  The lengths of the sulfur-gold bonds calculated
by Sellers {\em et al.} are used in this paper.

\begin{figure}[ht]
\includegraphics[bb=10 10 600 775 ,width=0.75\textwidth,clip]{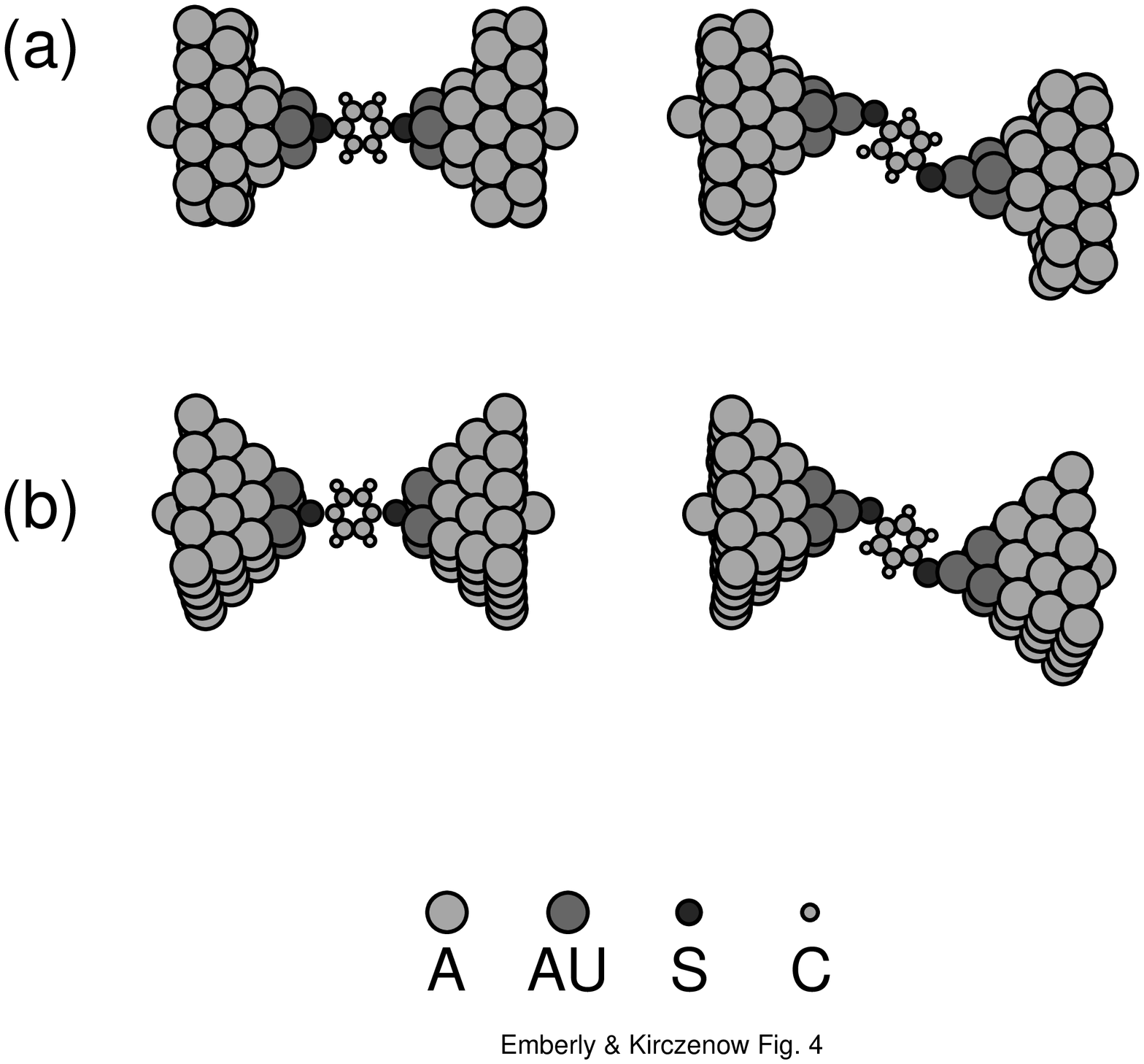}
\caption{Atomic diagrams of CMC systems.  (a) (111) Au gold
clusters with BDT.  (b) (100) Au gold clusters with BDT.  The
left figures correspond to hollow-site binding, with the right
figures showing on-site binding.  The posistion of the single
mode leads is shown in the diagrams by the single gold atom
attached to the outer faces of each of the gold clusters.  For the multimode
calculations, the outer two layers of each gold cluster form
the unit cell used for the leads.}
\label{fig:4}
\end{figure}

The cluster contact geometries to be considered are shown in
Fig. \ref{fig:4}.  In diagram (a), the gold clusters are
arranged in an ideal fcc (111) configuration, where the lattice
constant of bulk gold has been used to determine the spacing.
The left CMC system is for hollow site binding, where the S is
bonded over a triangle of gold atoms.  The perpendicular
distance of the S over the triangle is 1.9
Angstroms\cite{Sell93}.  The right diagram of
Fig. \ref{fig:4}(a) is for binding directly over a gold atom.
In this binding scheme the Au-S-C bond angle is not $180^{o}$.
For alkanethiolate it was found to be around
$110^{o}$\cite{Sell93}, but in order to match the experimental
lead-lead distance of 8.5 Angstroms for BDT\cite{Reed96}, we
chose it here to be $130^{o}$.  The bond length of the Au-S
bond was taken to be 2.35 Angstroms\cite{Sell93}.  For the
hollow site binding there are 50 Au atoms in each cluster and
51 in each for the on-site binding.

The cluster geometry shown in Fig. \ref{fig:4}(b) is for gold
(100).  Again the spacing has been determined using the
lattice parameter of bulk gold.  In the left figure the
sulfur is bonded over a hollow site.  In this case the
hollow site is defined by a square of gold atoms.  The
perpendicular distance of the S over this square was taken to
be 2.0 Angstroms\cite{Sell93}.  The right diagram is for the
case of binding directly over a gold atom.  Again the bond
angle was assumed to be $130^{o}$, but the bond length was now
assumed to be 2.15 Angstroms\cite{Sell93}.  The clusters for
hollow site binding have 54 Au atoms each and for on-site
binding 55 Au atom clusters are used.

In the last CMC geometry studied we took the left cluster to
be gold (100) and the right cluster to be (111).  This was
done to examine the effects of asymmetric binding, which may
well occur since the two nanocontacts need not be identical.
Both on-site binding and hollow site binding configurations
were considered.  All the bond distances and angles were based
on the previous geometries.

With these more realistic models of the contact region the
nature of the bonding of the clusters to the molecule is
clarified by considering the electronic structure of the CMC.
Extended H\"{u}ckel was used to calculate the energy levels of
the CMC.  The ideal leads and most of the atoms in the gold
cluster were modeled with just 6s orbitals.  However, the gold
atoms adjacent to the sulfur and responsible for most of the
bonding to the sulfur were treated using the full valence
orbital set.  For gold the electron Fermi level lies in the 6s
band so that this treatment of the electronic structure
makes it possible to study transport through fairly large
clusters, while at the same time retaining a realistic
description of the bonding between the BDT molecule and the
gold.

\begin{figure}[ht]
\includegraphics[bb= 40 10 475 775, width=0.75\textwidth,clip]{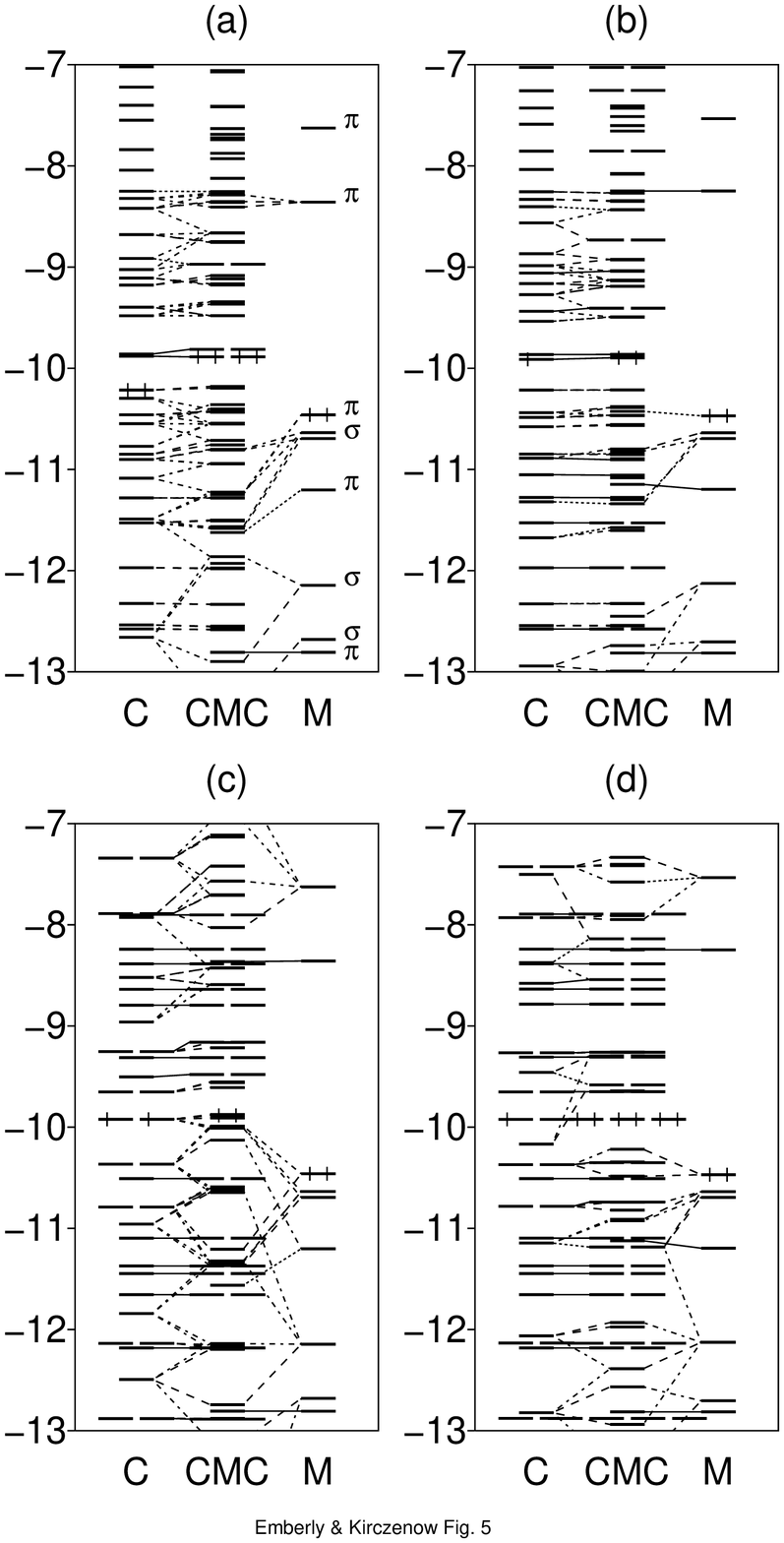}
\caption{Energy level diagrams for CMC systems.  (a) (111) Au
clusters with hollow-site binding.  (b) (111) Au clusters with
on-site binding.  (c) (100) Au clusters with hollow-site
binding.  (d) (100) Au clusters with on-site binding.  In the
energy level diagrams, C labels the energy levels of the
uncoupled left cluster, M labels the energy levels for the
free molecule, and CMC labels the energy levels for the bonded
clusters+molecule system.  Energies are in eV.}
\label{fig:5}
\end{figure}

The energy level diagrams for the various systems are shown in
Fig. \ref{fig:5}.  To clarify the nature of the various levels
the CMC was separated into three fragments: the left cluster
(denoted by C in the figure), BDT (denoted by M), and the right
cluster.  It is shown in the diagram how the levels of these
individual fragments get mixed into the overall electronic
structure.  In the binding geometries chosen above the coupling
of the molecule to the clusters is strong and the levels of the
isolated BDT are strongly affected.  It should be noted that in
these calculations the first sites for the single channel leads
were also included in the calculation since, as was seen
previously, the coupling $\bf{W}$ of the CMC to the ideal leads
also influences the electronic structure.  (In
Fig. \ref{fig:4}, the single gold atom bonded to the left and
right gold clusters of each diagram indicates where this 1D
gold chain was attached for the single mode calculations).  For
clusters of the size considered here this effect is not large,
but for smaller clusters the effect can distort the electronic
structure of the CMC very significantly.

Based on the four energy level schemes in the diagram it is
possible to make some general statements.  With the chosen
sizes of the clusters the level structure of the cluster is
quite dense.  As the cluster size increases this level
structure would eventually become the continuous energy bands
of bulk gold.  This is an attempt to take into account the
continuum nature of the spectrum of the gold contacts in the
model while still being able to address the chemical binding of
the contacts to the molecule.  It is also a useful way to
estimate the Fermi level of the leads by using the HOMO of the
cluster as an approximation.  In all of the diagrams it is seen
that the Fermi level of the clusters lies within the HOMO-LUMO
gap of the BDT.  It is also closer to the HOMO than to the
LUMO.  This has been suggested to be
true\cite{Reed96,Datta97_2} and the present calculations
support this idea.

Another significant aspect of the diagrams is the strong
mixing of the molecular states with cluster states.  This
mixing is significant in all diagrams.  For the clusters in
the (111) orientation, the molecular levels around the HOMO
are predominantly shifted downwards, whereas for the (100)
clusters the HOMO mixing is more symmetric in energy.  The
LUMO levels all tend to be mixed into the higher energy levels
of the CMC system.  The character of the levels also becomes
mixed; it can be seen that some $\pi$ levels of the molecule
are mixed with $\sigma$ states.  In all the diagrams there is
also a series of levels between -10 eV and -9 eV for which no
molecular content is shown explicitly.  However these states
contain complex, but small admixtures of many BDT levels.
From these energy diagrams it appears the BDT HOMO-LUMO are
not clearly identifiable in the energy level structure of the
combined system for these cases which involve strong binding
of the molecule to the clusters.  Thus transport in these
systems will be very different from that through a free
molecule.

The energy level diagrams for the asymmetric binding are not
shown since they are qualitatively similar to those in
Fig. \ref{fig:5} with the main difference being that
degeneracies in the energy level spectrum are broken due to
the lower symmetry of the system.

\section{Transmission and Conductance of CMC Systems with
Single Mode Leads}
The electronic transmission through the above systems will now
be calculated using leads with just a single propagating mode.
By using leads with just one energy band we will be able to
study the role of the chemistry in transport.  We will use the
single electron channel to probe the energy level structure of
the CMC systems.  The leads are constructed out of a unit cell
containing one gold atom with a 6s orbital.  The transmission
probability $T(E)$ was calculated for energies around the Fermi
level of the Au leads.  The coupling matrix, $\bf{W}$, was
determined by evaluating the H\"{u}ckel matrix elements between
the ideal leads' 6s orbital on the sites adjacent to the CMC
(shown as the single gold atoms attached to the outer faces of the gold 
clusters in Fig. \ref{fig:4}) and all the orbitals of the CMC.  The
Landauer formula was then used to obtain the current vs bias
voltage curve for each system.  The differential conductance
was calculated from this data.

\begin{figure}[ht]
\includegraphics[bb= 0 10 590 775 ,width=0.75\textwidth,clip]{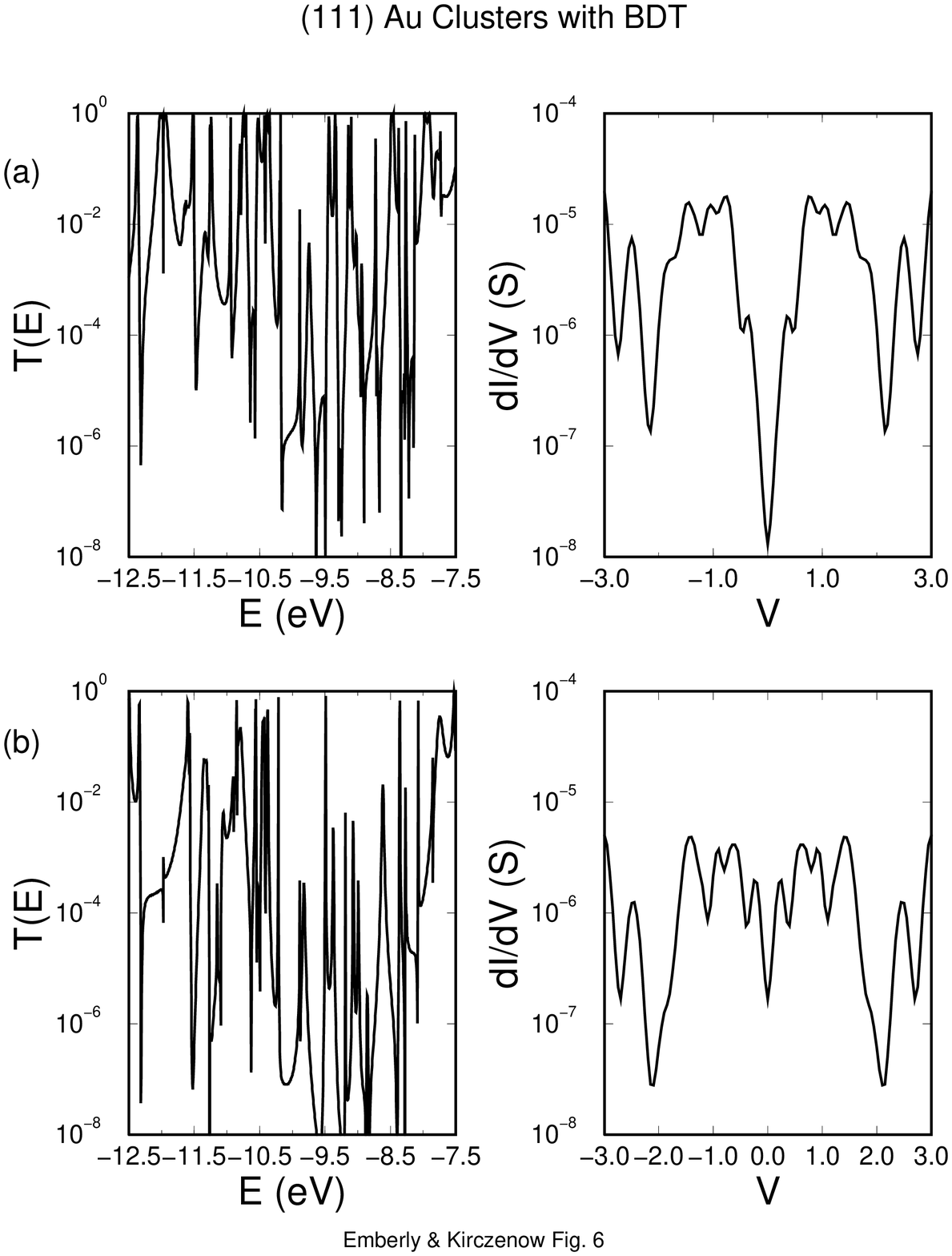}
\caption{Transmission and differential conductance for (111)
Au clusters with BDT.  (a) hollow-site binding.  (b) on-site
binding.}
\label{fig:6}
\end{figure}

In Fig. \ref{fig:6}, the transmission and differential
conductance diagrams are shown for the clusters in the (111)
orientation.  The top diagrams are for hollow-site binding
and the bottom for on-site binding.  There is significant
transmission throughout most of the energy range with some
overall reduction in the region from -10.5 eV to -9.5 eV.
This corresponds to the levels in the energy diagram that have
less molecular content.  The transmission for on-site binding
is on the whole of lesser magnitude than for the hollow-site
case.  This is because the molecule is less strongly bonded
and the number of channels entering the molecule is less since
it is only attached to one gold atom.  The transmitting states
are attributed to either delocalized $\pi$ or $\sigma$ bonding
states of the molecule.  For instance, in the transmission
diagram of Fig. \ref{fig:6}a, the resonances between -12.5 eV
and -11.5 eV are due to $\sigma$ states, whereas the states
around -10.5 eV to -11.5 eV are primarily $\pi$.  The
resonances between -8.5 eV and -8.0 eV are directly connected
with the molecules LUMO as seen from the energy diagram
Fig. \ref{fig:5}a.  The same can be said for the case of
on-site binding.  There however the mixing is not as
significant and the transport through states connected with
the HOMO and LUMO is more noticeable.  This is evidenced by
the resonances at -10.5 eV where the states are from the HOMO.
The states at -8.5 to -7.5 eV are from the two BDT states
around the LUMO. In both cases the mixing of the molecule has
led to the large transmission throughout the energy range.

The differential conductance curves for the two (111) clusters
are shown also.  The Fermi energy was arbitrarily chosen to be
-10.1 eV for both of these calculations.  Choosing a different
Fermi energy can result in significantly different conductance
characteristics for some calculations where the transmission
has fewer resonances.  However in the present case, because of the many 
resonances 
that occur in the energy range of interest, the conductance is relatively 
flat and not very
sensitive overall to the choice of Fermi energy.  It is also significantly 
higher (by two orders of magnitude) than what was found
experimentally\cite{Reed96}.  This has also been found in other
calculations on BDT\cite{Kemp97}.

\begin{figure}[ht]
\includegraphics[bb= 0 10 590 775 ,width=0.75\textwidth,clip]{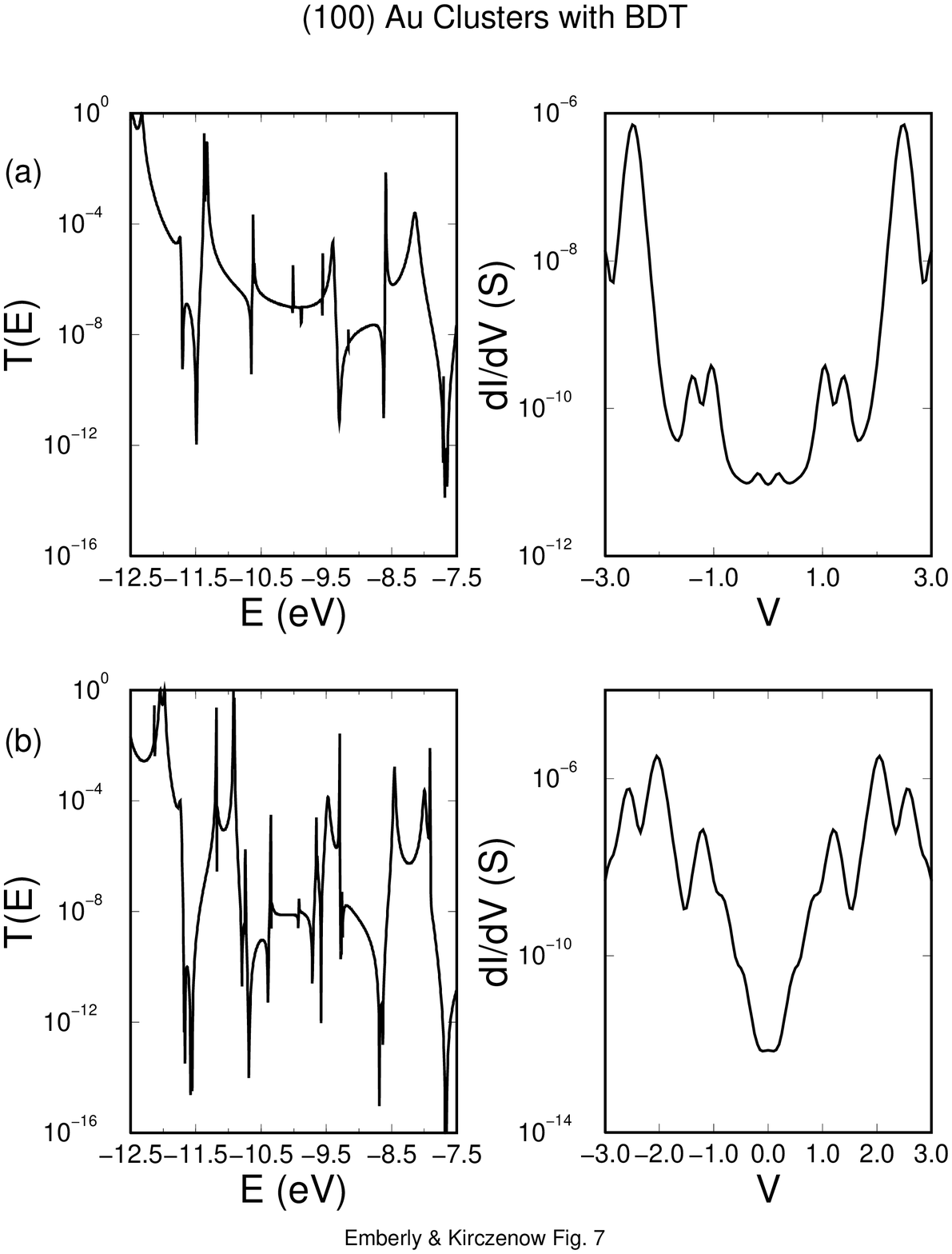}
\caption{Transmission and differential conductance for (100)
Au clusters with BDT.  (a) hollow-site binding.  (b) on-site
binding.}
\label{fig:7}
\end{figure}

The transmission curves for the (100) oriented clusters
presented in Fig. \ref{fig:7} have fewer resonances and the
overall magnitude is lower.  For this case the on-site
transmission (b) has more resonances than the hollow-site case
(a).  This is because the sulfur is bonded more strongly to the
single gold atom than over the four gold atoms.  Again the
molecular states are mixed with the cluster states.  But it can
be seen in the energy level diagrams, Fig. \ref{fig:5}(c,d),
that within the CMC spectrum there are levels that are almost
entirely cluster levels.  This is different than the (111) case
where only a few of the CMC states were entirely cluster
states.  These states transmit very poorly.  Those states that
are more transmissive can again be identified as either
delocalized $\sigma$ or $\pi$ states from the molecule.  In
both diagrams the resonances below -11 eV are due mostly to
$\sigma$ states, while the resonances around -9.5 eV to -10.5
eV are due to $\pi$ states and again the resonances around -8.5
eV have BDT LUMO content.  The lower magnitude in the
transmission curves results in lower conductance compared to
the previous results.  For the differential conductance
calculations the Fermi Energy was arbitrarily chosen to be
-10.0 eV.

\begin{figure}[ht]
\includegraphics[bb= 0 10 590 775 ,width=0.75\textwidth,clip]{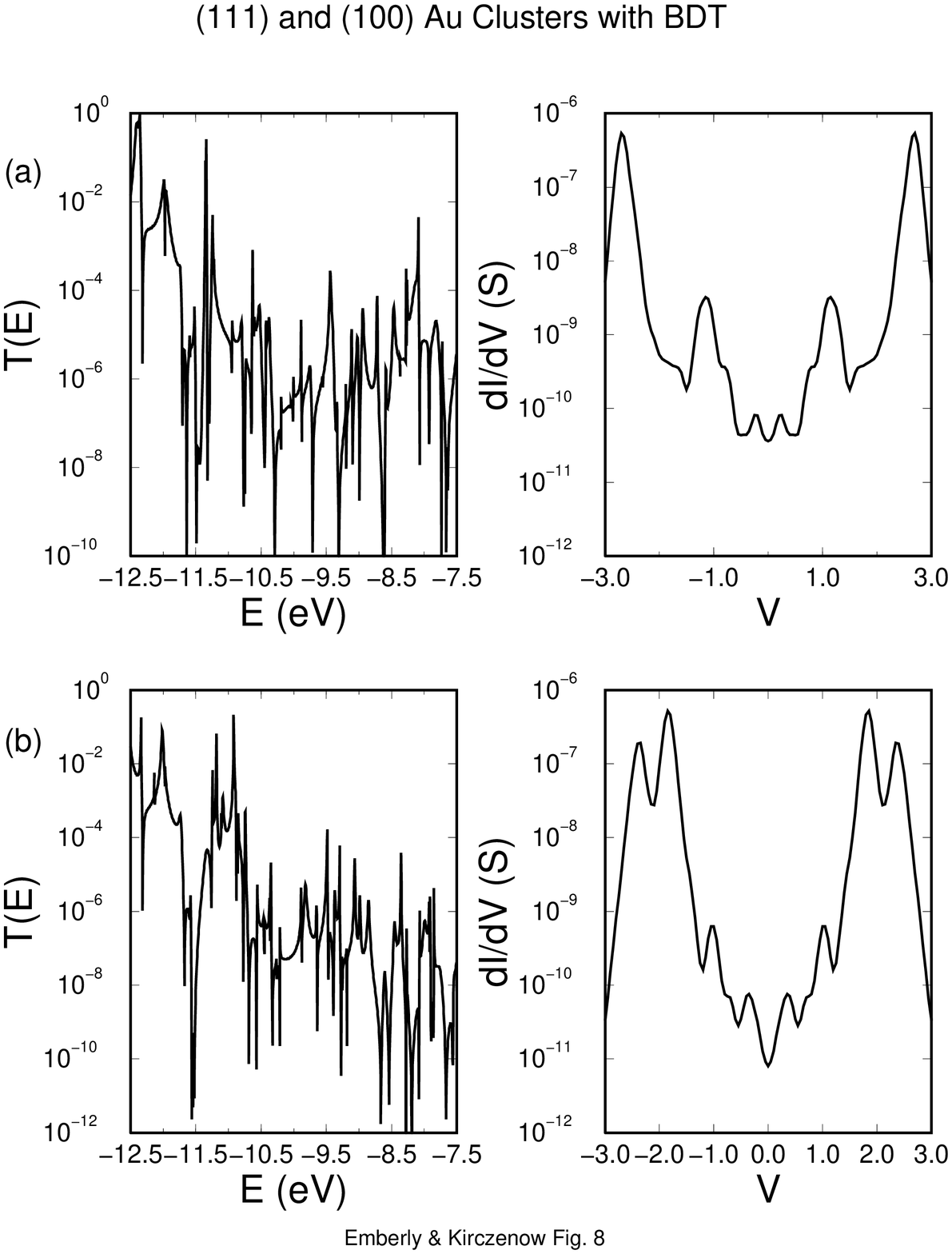}
\caption{Transmission and differential conductance for BDT
bonded to (100) and (111) Au clusters.  (a) hollow-site
binding.  (b) on-site binding.}
\label{fig:8}
\end{figure}

Finally, in Fig. \ref{fig:8} we present results for the BDT
molecule binding to two dissimilar clusters.  In the calculated
transmission curves, Fig. \ref{fig:8}(a,b), the transmission
now seems to share features of both the (111) and (100)
oriented clusters. The magnitude of the transmission is lower
than for the BDT connected to two (111) clusters
(Fig. \ref{fig:6}) and comparable to that for BDT connected to
two (100) clusters (Fig. \ref{fig:7}). The number of resonances
present is similar to the pure (111) case.  The differential
conductance curves were calculated for a Fermi energy of -10
eV.  Qualitatively, the structure of the curves agrees with the
experimental findings\cite{Reed96} in that the differential
conductance is very small at low bias and begins to increase
strongly at bias voltages of the order of 1 Volt for this
choice of Fermi energy. However, as for the cases of
binding between identical clusters discussed above the
differential conductance is again significantly larger than was
reported experimentally.

\section{Transmission and Conductance of CMC Systems
with Multi-mode Leads}
The previous section examined the transmission through CMC
systems that had 1-band ideal leads attached.  This allowed
the opportunity to relate resonances to the electronic
structure of the contact area.  We now attach multi-mode or
multi-channel leads to the CMC.  The term multi-channel lead is
used to refer to leads that have more than one energy band.  We
construct such leads out of unit cells that contain more than
one atomic orbital.  By considering leads with unit cells
containing more gold atoms we are able to model the many
energy modes that exist within macroscopic gold contacts.

We again consider two different geometric configurations for
the leads: (111) and (100) oriented gold contacts.  For the
lead in the (100) direction the unit cell is taken to consist
of two layers of gold atoms.  One layer has 16 atoms and the
other has 25.  Referring to the CMC diagrams previously shown,
the unit cells for the (100) left and right leads are the
outermost two layers of the gold clusters on the left and right
in Fig. \ref{fig:4}(a,b).  For the (111) oriented lead the unit
cell is again chosen to consist of 2 layers of gold atoms.  It
consists of two layers of 20 atoms each.  To construct a lead
with the structure of bulk gold in the (111) direction,
actually requires a unit cell containing three layers, but two
were chosen in this work to keep the number of modes similar
between the two orientations.  Again, the outermost two layers
of the gold clusters on either side in Fig. \ref{fig:4}(a,b)
comprise the left and right unit cells that we use.  These unit cells
are periodically repeated in 1D to form the ideal leads now used for
these multimode calculations.  We
use these unit cells with the gold atoms treated with just 6s
orbitals to calculate the band structure of the 1D leads.

For these multimode calculations the interaction matrix,
${\bf{W}}$, is taken to be the H\"uckel matrix elements between
all of the 6s orbitals in the unit cells directly adjacent to the
now smaller CMC, and all orbitals on the CMC.

\begin{figure}[ht]
\includegraphics[bb= 0 10 590 775 ,width=0.75\textwidth,clip]{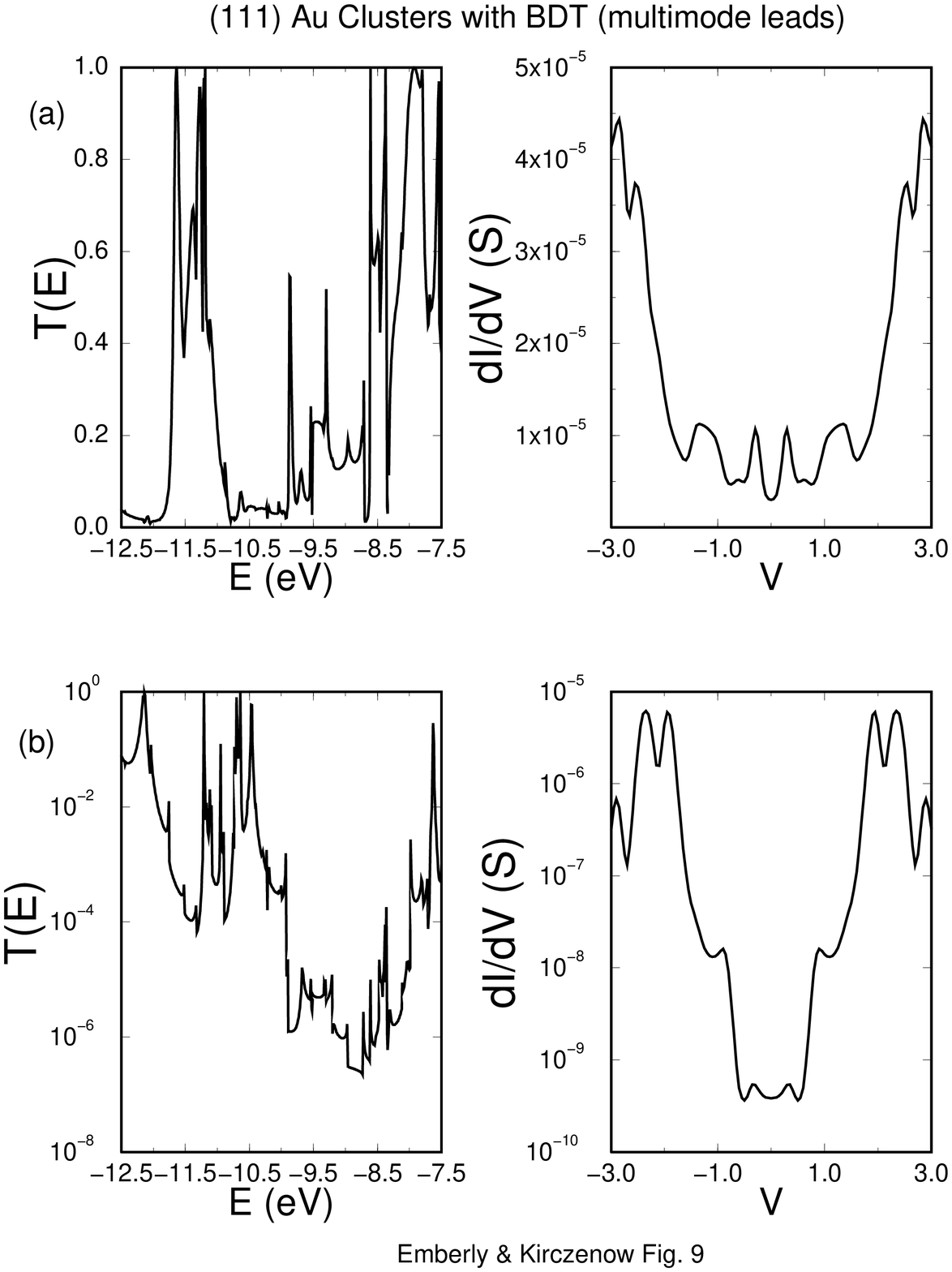}
\caption{Transmission and differential conductance for (111)
Au multi-mode leads with hollow site binding to BDT.  (a) is for
strong bonding to the BDT, while (b) is for weak bonding.}
\label{fig:10}
\end{figure}

The first case that is considered is that of hollow site
binding for the (111) leads.  The transmission diagram is shown
in Fig. \ref{fig:10}a.  The overall magnitude is up
significantly compared to the ideal 1D lead calculation.  The
peaks are also broadened.  There is strong transmission in the
energy regions where the cluster states have mixed with the
molecular states.  This occurs most prominently around -11.5
eV, where there exist resonances that can be connected with the
HOMO states of the molecule.  The other region of significant
transmission is at around -8 eV, which is due to CMC states
connected with the LUMO of the BDT.  The region in between has
resonances that arise from those states that are complex
admixtures of cluster with molecular levels.  The differential
conductance was calculated with a Fermi level chosen at -10 eV.
It is seen that the conductance magnitude is now much greater
than what it was for the ideal lead calculations.  The molecule
seems to be very conductive when attached strongly to the (111)
oriented wide leads.

\begin{figure}[ht]
\includegraphics[bb= 0 10 590 775 ,width=0.75\textwidth,clip]{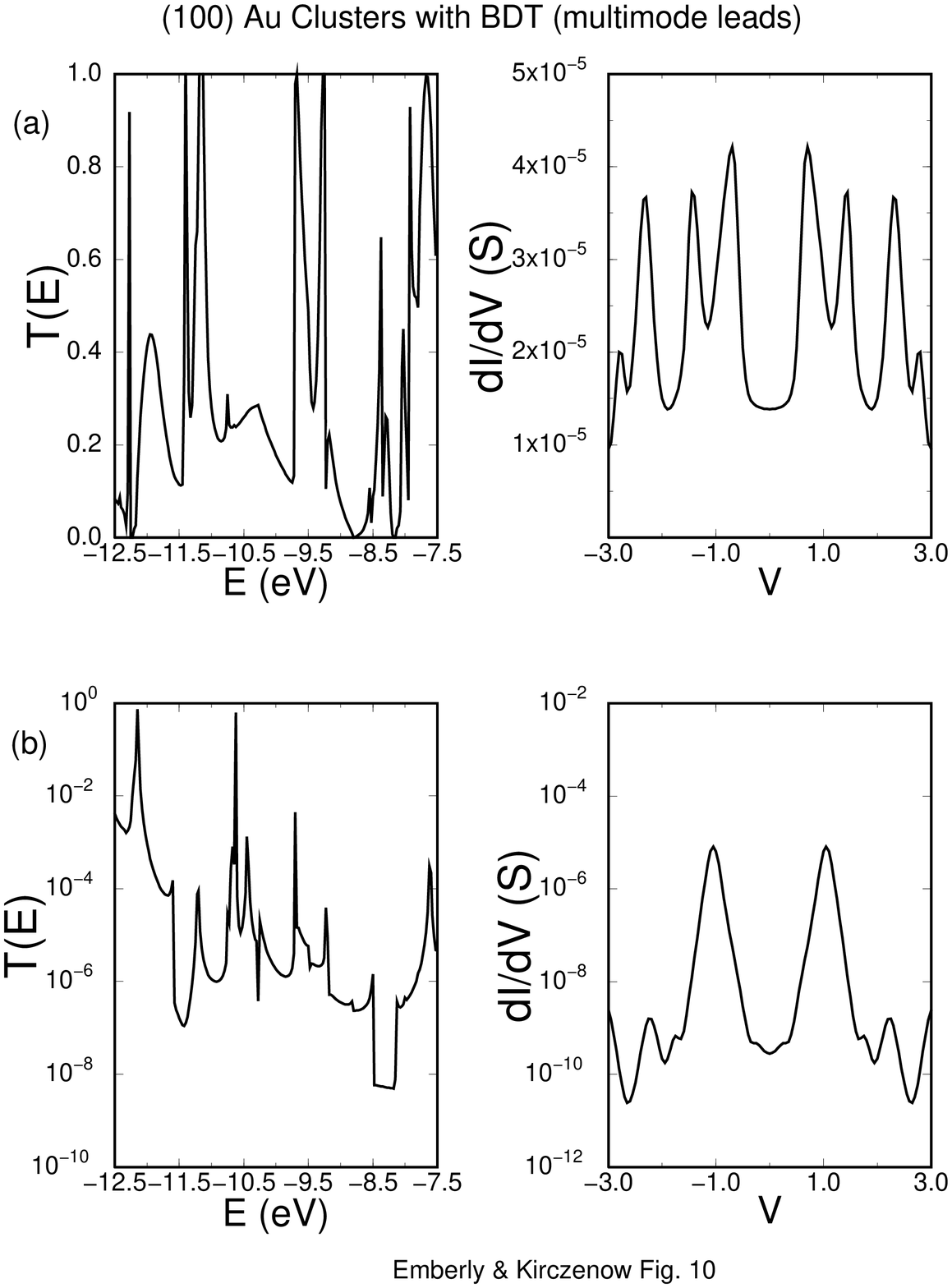}
\caption{Transmission and differential conductance for (100)
Au multi-mode leads with hollow site binding to BDT.  (a) is for
strong bonding to the BDT, while (b) is for weak bonding.}
\label{fig:11}
\end{figure}

The same general statements can be made about the transmission
through the molecule when attached to the leads oriented in
the (100) direction, Fig. \ref{fig:11}a.  Comparing the
transmission diagrams between the ideal 1D lead calculation
and wide lead calculation it is seen that the overall
structure is consistent.  However, as was the case above, the
magnitude is significantly up as well as the peaks having
broadened.  This again leads to a conductance which is greater
than that found for the ideal lead calculations.  The Fermi
energy was chosen again to be -10.0 eV.

The magnitude of the conductance found in the above two
calculations exceeds that found experimentally.  How far do
the leads need to be separated from the molecule in this model
to achieve experimental conductance magnitudes?  This question
is addressed by calculating the transmission for a system
consisting of (111) or (100) oriented wide leads and the
molecule with the perpendicular distance between the gold lead
and the sulfur atom being 3.5 Angstroms.

\begin{figure}[ht]
\includegraphics[bb= 0 10 590 775 ,width=0.75\textwidth,clip]{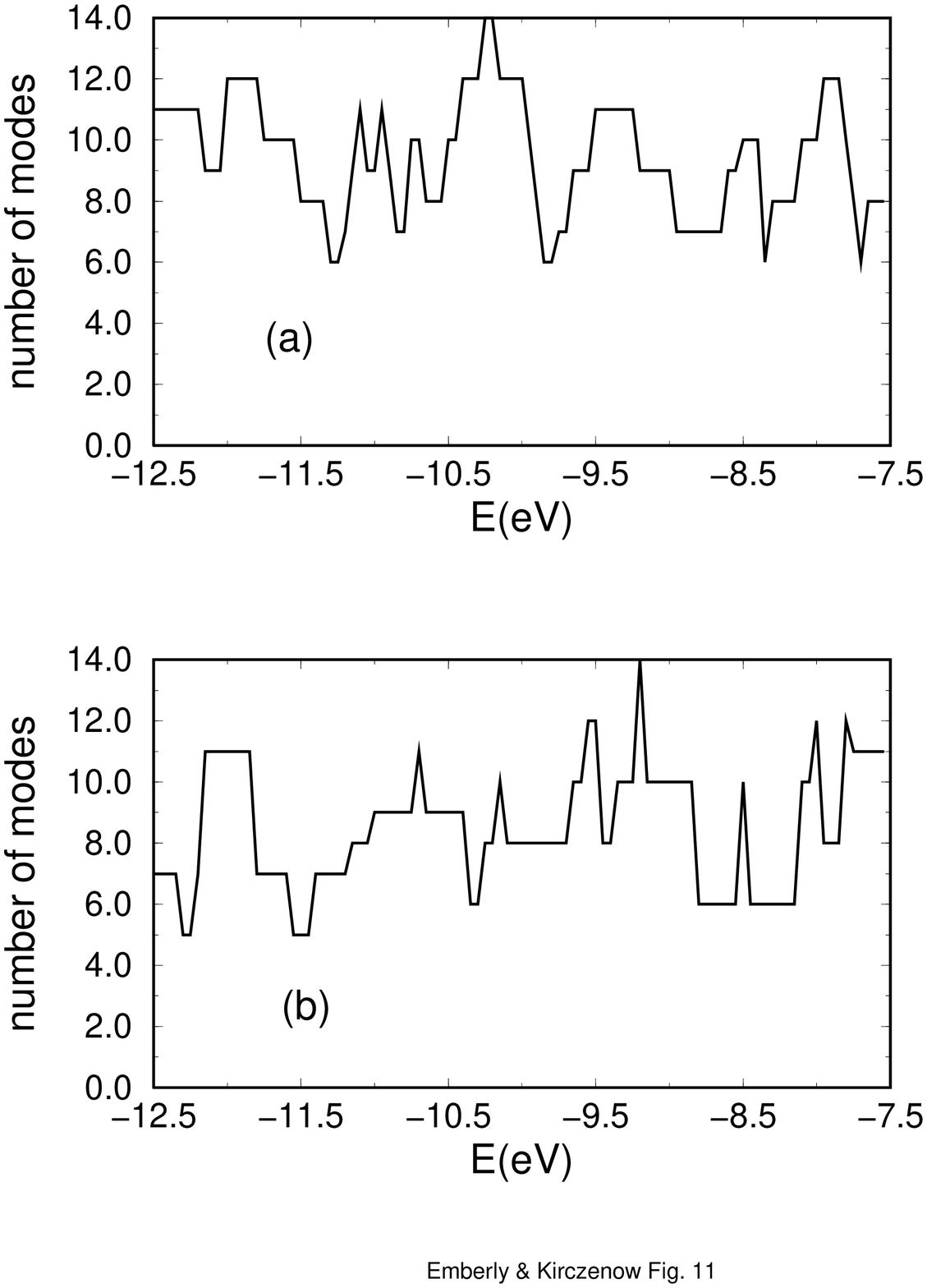}
\caption{Number of rightward propagating modes in 1d leads.
(a) is for (111) Au leads, and (b) is for (100) Au leads.}
\label{fig:12}
\end{figure}

The transmission and conductance diagrams for these weak
bonding cases are shown in Fig. \ref{fig:10}b and
Fig. \ref{fig:11}b respectively.  It can be seen that the
transmission has gone down significantly in the region between
-10.5 and -8 eV.  For both cases, there are also now strong
resonances that correspond almost exactly with the isolated
molecular levels.  (This is what we found previously for a
weakly coupled system).  The other peaks in the spectrum,
especially those in the region between -10.5 and -8 eV, can be
directly attributed to the number of modes propagating in the
lead.  In Fig. \ref{fig:12}, the number of modes propagating in
the leads as a function of energy is shown.  At these energies
the tunneling transmission will be proportional to the number
of modes in the lead.  For the (111) leads, the transmission
falls off sharply at -10.0 eV, corresponding to the drop in the
number of modes.  It then has peaks at around -9.5 and -8.5 eV,
where the number of modes increases.  Similar comments can be
made about the transmission for the (100) leads.  The weaker
coupling of the molecule to the contacts has lowered the
transmission and the conductance.  Because there are order
of magnitude fluctuation in $T(E)$ the conductance is now quite
sensitive to the selection of Fermi energy.  Different
conductance curves can be obtained if the Fermi energy is
chosen to occur on or off resonace.  The same Fermi energies
were used as in the strong bonding calcuations.  However, the
calculated conductance is still somewhat larger than that
reported experimentally, even for this case where in reality
the molecule would probably not bond to the leads for the bond
lengths considered.

\section{Conclusions}
In this paper we have presented a model which brings together
techniques from chemistry and physics to analyze electron
transport in molecular wires.  The coupling of the molecule to
the metallic contacts was studied by treating the contact
region as a single chemical entity.  By considering different
cluster geometries and binding mechanisms we have been able
to address two major unknowns in the experimental system:
contact geometry and the S-Au bonding.  We then proceeded to
study the transport properties of these systems by attaching
both single and multi-mode ideal leads.

We presented three different transmission calculations.  The
first calculation was on BDT bonded directly to 1D ideal
leads.  This simple system showed how for weakly coupled
systems the transmission is dominated by the energy levels of
the free molecule.  However, for strong coupling the energy
levels of the free molecule are distorted and the
transmission reflects this.  The second set of calculations,
which involved the attachment of different gold clusters to
the BDT, explored the role of CMC chemistry by using single
mode ideal leads.  The last calculation involved attaching
multi-mode ideal leads, which model more realistic metallic
contacts.

The second set of calculations showed that the
transmission is sensitive to the choice of both geometry and
bond.  In these strongly coupled systems we have seen that
the BDT states get mixed with the cluster states.  The
molecular HOMO and LUMO are not as prominent in the overall
CMC level structure.  The resulting transmission through the
strongly coupled system departs from that of the isolated
molecule.  It was also found that cluster geometries which
allow more cluster states to interact with the molecule,
lead to greater transmission.  This can be said of the (111)
clusters compared to the (100) clusters.  Lastly, the effect
of lead asymmetry was to reduce the overall transmission and
conductance relative to the purely (111) case.

The multi-mode lead calculations have shown that for the
present class of models based on one-electron tight binding
theory and the standard extended H\"{u}ckel parameters of
quantum chemistry, the BDT molecule is quite conductive for
reasonable bonding distances. Although the calculated
differential conductance displays a gap of the order of 1 Volt
at low bias in qualitative agreement with experiment, it was
necessary to make the bond length between the molecule and the
gold leads unphysically large in order to reduce the overall
transmission magnitude to the experimentally observed
range. The reason for this quantitative discrepancy between
the theory and experiment is not known at present.

While transport through molecular wires has been been
discussed theoretically for some time, experiments on
conduction through a single molecule chemically bonded between
metallic contacts have only just begun to be performed and to
provide a testing ground for theories. Clearly much remains to
be learned both experimentally and theoretically about the
mechanisms of electronic transport through these novel
devices.

\section{Acknowledgements}
We would like to thank Ross Hill, Christian Joachim, Mathieu
Kemp, Mark Ratner, and Mark Reed for helpful discussions.
For the extended H\"{u}ckel calculations, we would like to
acknowledge the EHC program, Yaehmop, by Greg Landrum.  This
work was supported by NSERC.

\section{Appendix}
The modes for the wide leads were calculated using standard
transfer matrix theory.  This method yields all the reduced
wavenumbers, $y^{\alpha}$, and the Bloch coefficients
$c^{\alpha}_j$, for all the modes with energy $E$ in a lead.

The transfer matrix considered here connects a Bloch state,
$|\Psi\rangle = \sum c_j e^{i n y} |n, j\rangle = \sum
\Psi_{n,j} |n,j\rangle$, between nearest neighbour sites.
This can be written as
\begin{equation}
\left( {\bf{T}}(E) \right) \left( \begin{array}{c}
{\bf{\Psi_n}} \\ {\bf{\Psi_{n-1}}} \end{array} \right)
= \left( \begin{array}{c} {\bf{\Psi_{n+1}}} \\
{\bf{\Psi_n}}
\end{array} \right)
\end{equation}

For a 1D chain treated in the tight-binding approximation,
each unit cell will have a matrix $\epsilon$ containing the
site energies and intra-cell hopping energies and a matrix
$\beta$ containing the inter-cell hopping energies.  The
transfer matrix for such a chain is
\begin{equation}
\left( \begin{array}{cc} \beta^{-1}\cdot(E{\bf{1-\epsilon}}) &
-\beta^{-1}\cdot\beta^T \\ \bf{1} & \bf{0}
\end{array} \right)
\end{equation}

The eigenvalues and eigenvectors of this matrix provide the
reduced wavenumbers and Bloch coefficients respectively.
%
%


\begin{references}
%
\bibitem{Avi74} A. Aviram and M. A. Ratner,
Chem. Phys. Lett. {\bf 29}, 257 (1974).

\bibitem{Fisch95} C. M. Fischer, M. Burghard, S. Roth and
K. von Klitzing, Surf. Sci. {\bf 361/362}, 905 (1995).

\bibitem{Zhou97} C. Zhou, M. R. Deshpande, and M. A. Reed,
Appl. Phys. Lett. {\bf 71}, 611 (1997).

\bibitem{Andres96} R. P. Andres, J. D. Bielefeld,
J. I. Henderson, D. B. Janes, V. R. Kolagunta, C. P. Kubiak,
W. J. Mahoney, and R. G. Osifchin, Science {\bf 273}, 1690
(1996).

\bibitem{Mirkin96} C. A. Mirkin, R. L. Letsinger, R. C. Mucic,
and J. J. Storhoff, Nature {\bf 382}, 607 (1996).

\bibitem{Bumm96} L. A. Bumm, J. J. Arnold, M. T. Cygan,
T. D. Dunbar, T. P. Burgin, L. Jones II, D. L. Allara,
J. M. Tour, P. S. Weiss, Science {\bf 271}, 1705 (1996).

\bibitem{Stipe97} B. C. Stipe, M. A. Rezaei, W. Ho, S. Gao,
M. Persson and B. I. Lundqvist, Phys. Rev. Lett. {\bf 78},
4410 (1997).

\bibitem{Datta97_2} S. Datta, W. Tian, S. Hong,
R. Reifenberger, J. I. Henderson and C. P. Kubiak,
Phys. Rev. Lett. {\bf 79}, 2530 (1997).

\bibitem{Reed96} M. A. Reed, C. Zhou, C. J. Muller,
T. P. Burgin, and J. M Tour, Science {\bf 278}, 252 (1997).

\bibitem{Samant96} M. P. Samanta, W. Tian, S. Datta,
J. I. Henderson, and C. P. Kubiak, Phys. Rev. B {\bf 53},
R7626 (1996).

\bibitem{Kemp96_1} M. Kemp, A. Roitberg, V.  Mujica, T. Wanta
and M. A. Ratner, J. Phys. Chem {\bf 100}, 8349 (1996).

\bibitem{Mago97} M. Magoga, and C. Joachim, Phys. Rev. B
{\bf 56}, 4722 (1997).

\bibitem{Datta97_1} S. Datta, and W. Tian, Phys. Rev. B {\bf
55}, R1914 (1997).

\bibitem{Ember98_1} E. Emberly and G. Kirczenow, Ann. New York
Academy of Science, {\bf{852}}, 54 (1998).

\bibitem{Mujic96} V. Mujica, M. Kemp, A.  Roitberg and
M. Ratner, J. Chem. Phys. {\bf 104}, 7296 (1996).

\bibitem{Joach96} C. Joachim, and J. F. Vinuesa,
Europhys. Lett. {\bf 33}, 635 (1996).

\bibitem{Lan57} R. Landauer, IBM J. Res. Dev. {\bf 1}, 223
(1957); R. Landauer, Phys. Lett. {\bf 85A}, 91 (1981)

\bibitem{Datta95} For a comprehensive review of Landauer
theory and electron transport see S. Datta,
{\em Electronic Transport in
Mesoscopic Systems}, Cambridge University Press, Cambridge,
1995.

\bibitem{Sautet88} P. Sautet and C. Joachim, Phys. Rev. B {\bf
38}, 12238 (1988).

\bibitem{Kemp96_2} M. Kemp, A. Roitberg, V. Mujica,
T. Wanta and M. Ratner, J. Phys. Chem. {\bf 100},
8349 (1996) and references therein.

\bibitem{Ember98_2} E. Emberly and G. Kirczenow, unpublished
(1998).

\bibitem{Laibin91} P. E. Laibinis, G. M. Whitesides,
D. L. Allara, Y. T. Tao, A. N. Parikh, R. G. Nuzzo,
J. Am. Chem. Soc. {\bf 113}, 7152 (1991).

\bibitem{Corr97} A. Correia, and N. Garcia, Phys. Rev. B
{\bf 55}, 6689 (1997).

\bibitem{Kondo97} Y. Kondo, and K. Takayanagi,
Phys. Rev. Lett. {\bf 79}, 3455 (1997).

\bibitem{Sell93} H. Sellers, A. Ulman, Y.  Schnidman and
J. E. Eilers, J. Am. Chem. Soc. {\bf 115}, 9389 (1993).

\bibitem{Beard98} K. M. Beardmore, J. D. Kress, N.
Gr{\o}nbech-Jensen, and A. R. Bishop,
Chem. Phys. Lett. {\bf{286}}, 40 (1998).

\bibitem{Kemp97} M. Kemp et al and S. Datta et al,
Molecular Electronics - Science and Technology Conference.
Unpublished (1997).

\end{references}
\end{document}